\documentclass[english]{article}
\usepackage[T1]{fontenc}
\usepackage[latin9]{inputenc}
\usepackage{textcomp}
\usepackage{amsmath}
\usepackage{graphicx}

\makeatletter
\usepackage{spconf,amsmath,graphicx}

\renewcommand{\author}{\name}

\address{$^{\star}$Radiology \& Biomedical Imaging and $^{\dagger}$Child Study Center, Yale School of Medicine, New Haven, CT\\
$^{\ast}$Biomedical Engineering and $^{\ddagger}$Electrical Engineering, Yale University, New Haven, CT}

\usepackage[font={small}]{caption}

\usepackage{enumitem}
\setlist{nolistsep}
\setlist[enumerate]{leftmargin=*}

\usepackage{fancyhdr}
\@ifundefined{showcaptionsetup}{}{%
 \PassOptionsToPackage{caption=false}{subfig}}
\usepackage{subfig}
\makeatother

\usepackage{babel}
\begin{document}

\title{Estimating Reproducible Functional Networks Associated with Task
Dynamics using Unsupervised LSTMs }

\author{Nicha C. Dvornek$^{\star\ast}$, Pamela Ventola$^{\dagger}$, and
James S. Duncan$^{\ast\star\ddagger}$\thanks{This work was supported by NIH grants R01MH100028 and R01NS035193.}}
\maketitle
\begin{abstract}
We propose a method for estimating more reproducible functional networks
that are more strongly associated with dynamic task activity by using
recurrent neural networks with long short term memory (LSTMs). The
LSTM model is trained in an unsupervised manner to learn to generate
the functional magnetic resonance imaging (fMRI) time-series data
in regions of interest. The learned functional networks can then be
used for further analysis, e.g., correlation analysis to determine
functional networks that are strongly associated with an fMRI task
paradigm. We test our approach and compare to other methods for decomposing
functional networks from fMRI activity on 2 related but separate datasets
that employ a biological motion perception task. We demonstrate that
the functional networks learned by the LSTM model are more strongly
associated with the task activity and dynamics compared to other approaches.
Furthermore, the patterns of network association are more closely
replicated across subjects within the same dataset as well as across
datasets. More reproducible functional networks are essential for
better characterizing the neural correlates of a target task. 

\end{abstract}

\section{Introduction}

The canonical approach to task-based functional magnetic resonance
image (fMRI) analysis has been mass univariate analysis at the voxel
level \cite{Monti2011}. While this allows for potentially high specificity
in locating brain regions that activate with task, the noisy fMRI
data lends to many false positives and the relationship between voxels
in different regions is ignored. Summarizing the voxel data by regions
of interest (ROIs) helps to alleviate the noisy signal problem, although
regressing task against individual ROIs again ignores the potential
relationships between distinct brain regions.

A multivariate approach that analyzes functional brain networks allows
for a higher systems-level view of neurocognitive functions that are
associated with a given task. In addition, considering several brain
regions in a network may better characterize the task-associated activity
and allow for a more robust representation of task-related brain changes. 

The functional brain activity can be decomposed into separate networks
using standard statistical tools such as principal component analysis
(PCA) and independent component analysis (ICA). However, a challenge
in determining functional networks, and more generally in fMRI analysis,
is the question of reproducibility. The number of subjects in a task
fMRI study is often smaller, making reproducibility of results challenging.
While traditional methods for finding functional networks are based
on analytical approaches that fit an entire dataset, predictive methods
that look to generalize well to new data may improve the reproducibility
of functional networks. 

In this paper, we propose to use recurrent neural networks with long
short term memory (LSTMs) \cite{Hochreiter1997} to estimate more
reproducible functional networks which are strongly associated with
task dynamics. Recently, LSTMs have been applied to fMRI data, e.g.,
to model the fMRI activity given a stimulus input \cite{Gueclue2017}
and for classification tasks based on fMRI \cite{Dvornek2019}. We
focus on using the LSTM's strength in signal generation, as demonstrated
in applications such as text generation \cite{Sutskever2014}. We
describe how the functional networks can be estimated with unsupervised
training of the LSTM model and then used for followup task-based analysis.
We show improved association of functional networks with task dynamics
and more reproducible results within and across 2 task fMRI datasets.

\section{Methods}

\subsection{LSTM for fMRI Signal Prediction}

The use of an LSTM-based network for fMRI data generation was recently
proposed by Dvornek et al.~\cite{Dvornek2019} for use as an auxiliary
task to enhance learning of a discriminative task. We adopt the basic
framework but focus solely on the unsupervised learning task of predicting
the fMRI time-series data for $N$ ROIs at time $x_{T+1}\in R^{N}$
given the time-series data from the previous $T$ time points $\left\{ x_{1},\ldots x_{T}\right\} $.
The ROI time-series data for $T$ time points are directly input into
an LSTM layer with $K$ units. The output of the LSTM layer, i.e.,
the hidden state $h_{T}\in R^{K}$, is then fed to a fully connected
layer with $N$ nodes, representing the $N$ ROI signals at time $T+1$.
Note that while the network is trained in a supervised manner, this
approach is truly unsupervised, as no labels or additional information
about the data is required. 

The functional networks will be represented by the $K$ units of the
LSTM. The network tries to learn the interaction between the $N$
individual ROIs and the $K$ functional networks by generating the
ROI time-series data from the decomposed functional networks,
\[
\widehat{x_{T+1}}=W_{f}h_{T}+b_{f},
\]
where $\widehat{x_{T+1}}$ is the predicted ROI signal at time $T+1$,
$W_{f}$ is a matrix of weights, and $b_{f}$ is a vector of biases.
The membership to a functional network $k$ is defined by the weights
in column $k$ of $W_{f}$. Different from \cite{Dvornek2019}, we
include both L1 regularization (controlled by a hyperparameter $l$)
and a non-negative constraint on the weights $W_{f}$. This encourages
functional networks to have sparse ROI membership and to work in a
cooperative manner to produce the ROI signals.\vspace{-2bp}

\subsection{Functional Networks Associated with Task Dynamics }

To determine functional networks that are associated with a given
task performed during the fMRI scan, we perform a group-wise analysis.
While any method can be applied (for example, a general linear model),
here we measure the correlation between the expected fMRI signal based
on the task design and the activity of the functional networks. First,
the task stimulus design is convolved with a canonical hemodynamic
response function. We also compute the temporal derivative of the
expected fMRI response. While all subjects in a dataset perform the
same task, the timing of each individual run contains very small variations.
Thus, we average the expected fMRI responses and their derivatives
across all subjects to obtain a mean ideal signal and its derivative,
which we refer to in the following as the task design signals. 

The activity of the functional networks is represented by the output
of the LSTM layer, $h_{t}$, i.e., $h_{t}\left(k\right)$ represents
the activity of functional network $k$ at time $t$. For each subject,
we extract the output of the LSTM for every time point $t\geq T$
(since the first $T$ time points are required for fMRI signal prediction).
We then average the LSTM outputs across all subjects to obtain the
mean activity of each functional network.

Finally, we compute the correlation between the mean expected fMRI
signal (from time $T$ to the end of the scan) and the mean activity
of each functional network, and the correlation between the mean temporal
derivative of the expected fMRI signal and the mean activity of each
functional network. A functional network with high correlation with
the mean expected fMRI signal is associated with the task activity
itself. A functional network with high correlation with the mean derivative
of the expected fMRI signal is associated with \emph{dynamic changes}
in the task. For example, in a task paradigm with block design, the
functional network would be associated with the switching between
2 conditions.

\section{Experiments}

\subsection{Data and Preprocessing}

Data were acquired for 2 separate datasets (originally for different
studies) using the same biological motion perception paradigm \cite{Kaiser2010},
protocol, and scanner. Each subject viewed a series of alternating
blocks of point-light displays of biological motion and scrambled
biological motion (\textasciitilde{}24 s/block, 6 of each condition).
Dataset 1 included 82 children with autism (age = $10.79\pm3.23$
years, IQ = $83.88\pm43.9$2) and 48 typical controls (age = $9.31\pm3.90$
years, IQ = $86.38\pm34.20$) matched for age and IQ. Dataset 2 included
21 children with autism (age = $6.05\pm1.24$ years, IQ = $102.00\pm17.87$)
and 19 typical controls (age = $6.42\pm1.29$ years, IQ = $111.47\pm14.45$),
again matched for age and IQ. Each subject underwent a BOLD fMRI scan
(TR = 2000 ms, TE = 25 ms, flip angle = 60\ensuremath{\circ}, voxel
size = 3.44\texttimes 3.44\texttimes 4mm\textsuperscript{3}) acquired
on a Siemens MAGNETOM Trio TIM 3T scanner. 

Images were preprocessed in FSL \cite{Jenkinson2012}, including motion
correction, interleaved slice timing correction, brain extraction,
4D mean intensity normalization, spatial smoothing (5 mm FWHM), data
denoising via ICA-AROMA \cite{Pruim2015}, nuisance regression using
white matter and cerebrospinal fluid, and high-pass temporal filtering
(100 s). Functional MRI were registered to the standard MNI brain
and parcellated into 90 cerebral brain regions using the AAL atlas
\cite{Tzourio-Mazoyera2002}. The mean time-series (146 and 156 time
points for Dataset 1 and 2, respectively) was extracted from each
ROI and standardized (subtracted the mean and divided by the standard
deviation).

To effectively train the LSTM models, we augmented the data for each
subject by extracting all possible windows of data with length $T=30$
(60 s of scan time). Thus, the LSTM network receives inputs of size
$30\times90$. Dataset 1 was augmented from 130 to a total of 15080
samples, while Dataset 2 was augmented from 40 to a total of 5040
samples.

\vspace{-1bp}

\subsection{\noindent Experimental Methods}

The proposed LSTM models for generating fMRI time-series data were
trained separately for each dataset. We implemented the models in
Keras using the mean squared error loss function, $l=0.0001$ for
the L1 weight regularization of $W_{f}$ for the fully connected layer,
the AMSGrad variant \cite{Reddi2018} of the Adam optimizer (learning
rate = 0.001), a batch size of 32, and 20 epochs of training. The
best epoch was chosen based on the minimum loss of a monitored validation
set. 

We compared the proposed LSTM method to 3 other approaches: 1) Using
the original ROIs. We considered each ``network'' to contain only
1 ROI, to investigate the potential advantages of truly functional
network approaches. 2) Using principal component analysis (PCA). The
ROI time-series data for all subjects was concatenated across time
and PCA was performed. The functional networks were then defined by
the principal components, and the activity of each network was given
by the score from projecting the data onto the principal components.
3) Using independent component analysis (ICA). Again, the data for
all subjects was concatenated across time. Group ICA \cite{Calhoun2009}
using the fastICA package for MATLAB \cite{fastICA} with default
parameters was performed. The functional networks were then defined
by the independent components, and the activity of each network was
given by the mixing matrix. For each of the network analysis approaches,
we tried 2 values for the number of functional networks $K$ (i.e.,
number of LSTM units or number of principal/independent components):
25 or 50. 

For model evaluation, we used 10-fold cross-validation of each dataset,
with 10\% for testing and 90\% for training. For the LSTM models,
10\% of the training data was withheld as validation data. Partitions
of the dataset were performed in a subjectwise manner, such that all
samples from the same subject are kept in the same partition. We computed
the correlations between estimated functional networks and the design
signals (task activity and task dynamics for the 2 task stimuli),
resulting in 4 correlation vectors with length $K$: $c_{BA}$ (biological
motion activity), $c_{BD}$ (biological motion dynamics), $c_{SA}$
(scrambled motion activity), and $c_{SD}$ (scrambled motion dynamics).
The correlations were computed for data in the training set, as well
as separately the test set.

We assessed the overall ability of the estimated functional networks
to capture task-relevant activity by the maximum and minimum correlations
for each design signal. Furthermore, we assessed the reliability and
reproducibility of the estimated functional networks in 2 ways. First,
to assess within dataset robustness, for each design signal we measured
the correlation of the corresponding computed measures between the
training and test sets (e.g., correlation between $c_{BA}$ computed
from training and test data). We expect that for a reproducible functional
network, the same correlation between that network's activity and
the task design signals should be observed in the training and test
sets. Second, to assess robustness across datasets, we took the functional
networks localized in Dataset 1 (2), extracted the corresponding activity
for these networks in Dataset 2 (1), and measured the maximum/minimum
correlations in the test sets. Reliable functional networks that capture
task-relevant activity should produce large correlations with the
task design signal in the independent dataset. Paired two-tailed t-tests
were used to compare the corresponding results from different methods
across cross-validation folds, with significance level $\alpha=0.05$.\vspace{-1bp}

\subsection{Results and Discussion}

\begin{figure}[t]
\noindent \begin{centering}
\negthinspace{}\negthinspace{}\subfloat[]{\includegraphics[width=0.36\columnwidth]{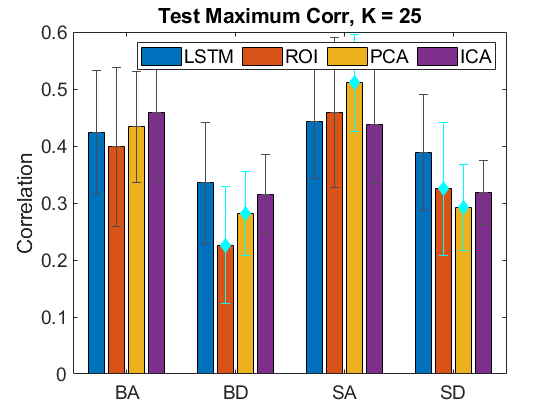}}\negthickspace{}\negthickspace{}\subfloat[]{\includegraphics[width=0.36\columnwidth]{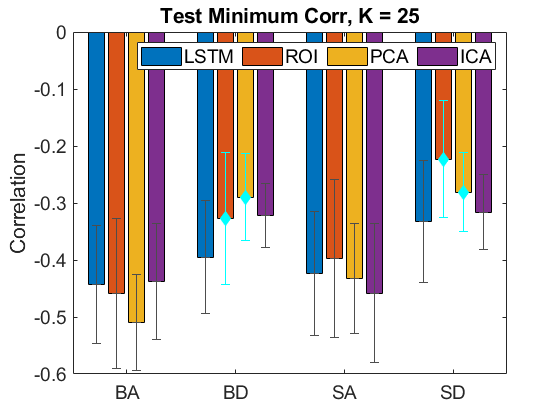}}\negthickspace{}\negthickspace{}\subfloat[]{\includegraphics[width=0.36\columnwidth]{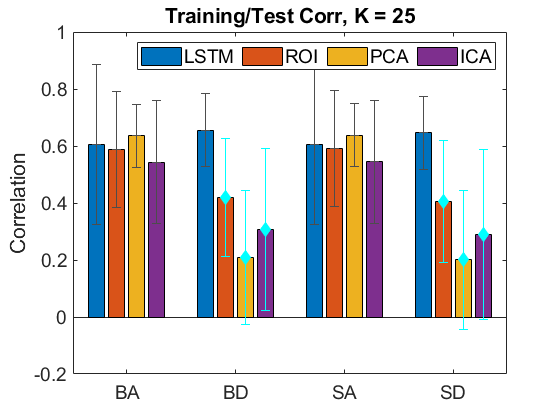}}
\par\end{centering}

\noindent \begin{centering}
\subfloat[]{\includegraphics[width=0.36\columnwidth]{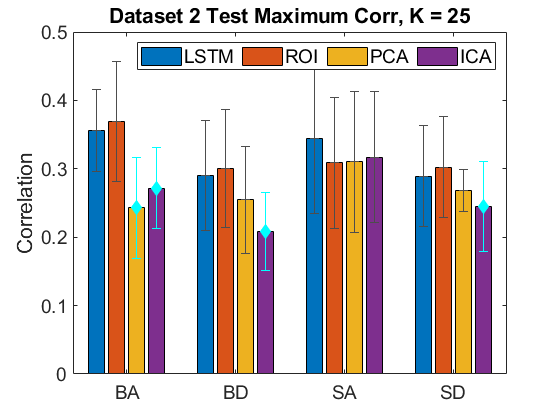}}\subfloat[]{\includegraphics[width=0.36\columnwidth]{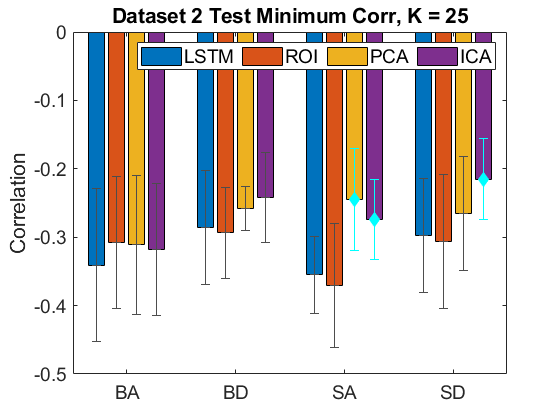}}
\par\end{centering}

\caption{\label{fig:dataset1_25}Results for Dataset 1, with $K=25$ networks.
(a) Maximum and (b) minimum correlation between functional network
activity and task design signals. (c) Correlation between patterns
of activity associated with task design signals from the training
and test set. (d) Maximum and (e) minimum correlation between functional
network activity in Dataset 2 using networks estimated by Dataset
1 and task design signals. BA = biological motion task activity, BD
= biological motion task dynamics, SA = scrambled motion task activity,
SD = scrambled motion task dynamics, LSTM = our method (blue), ROI
= individual ROI analysis (red), PCA = principal component analysis
(yellow), ICA = independent component analysis (purple). Cyan markers
denote statistically significantly different results compared to our
LSTM method (paired two-tailed t-test, $p<0.05$). }
\end{figure}
\begin{figure}[t]
\noindent \begin{centering}
\negthinspace{}\negthinspace{}\subfloat[]{\includegraphics[width=0.36\columnwidth]{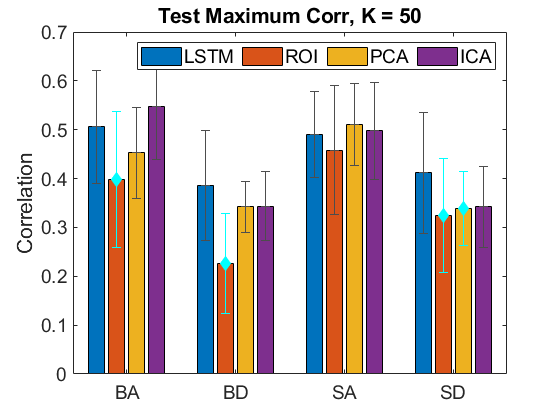}}\negthickspace{}\negthickspace{}\subfloat[]{\includegraphics[width=0.36\columnwidth]{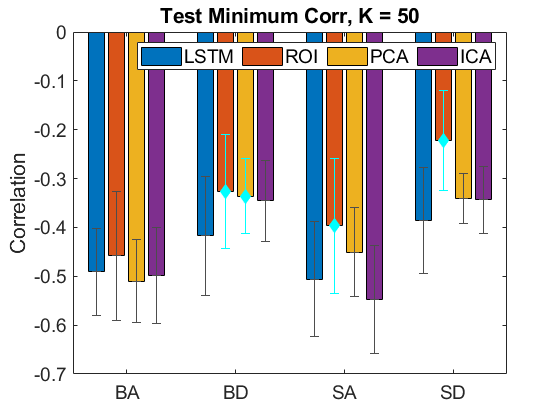}}\negthickspace{}\negthickspace{}\subfloat[]{\includegraphics[width=0.36\columnwidth]{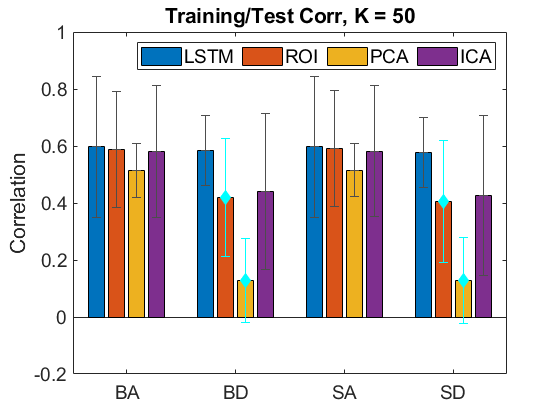}}
\par\end{centering}

\noindent \begin{centering}
\subfloat[]{\includegraphics[width=0.36\columnwidth]{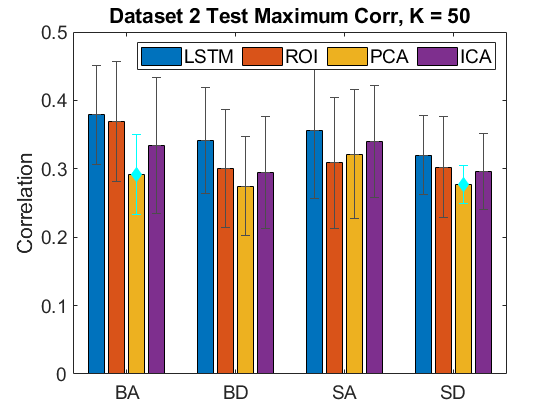}}\subfloat[]{\includegraphics[width=0.36\columnwidth]{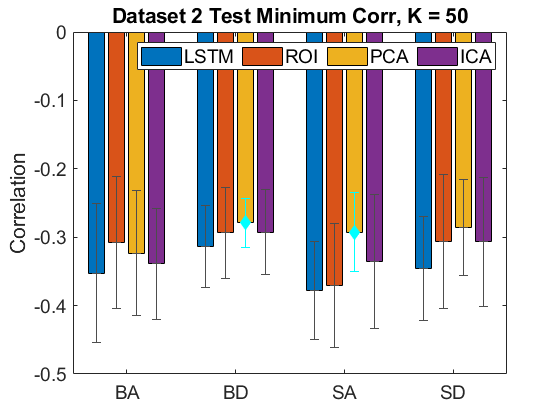}}
\par\end{centering}

\caption{\label{fig:dataset1_50}Results for Dataset 1, with $K=50$ networks.
See Fig.~1 caption for legend.}
\end{figure}
\begin{figure}
\noindent \begin{centering}
\negthinspace{}\negthinspace{}\subfloat[]{\includegraphics[width=0.36\columnwidth]{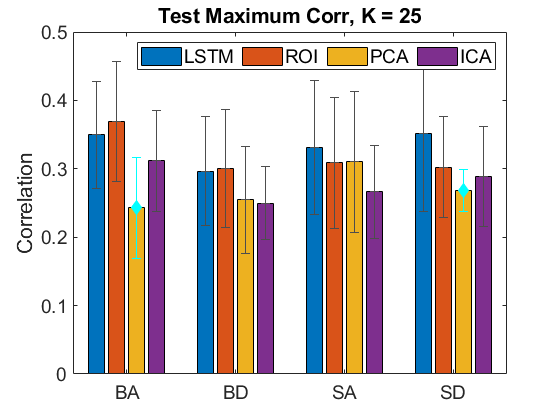}}\negthickspace{}\negthickspace{}\subfloat[]{\includegraphics[width=0.36\columnwidth]{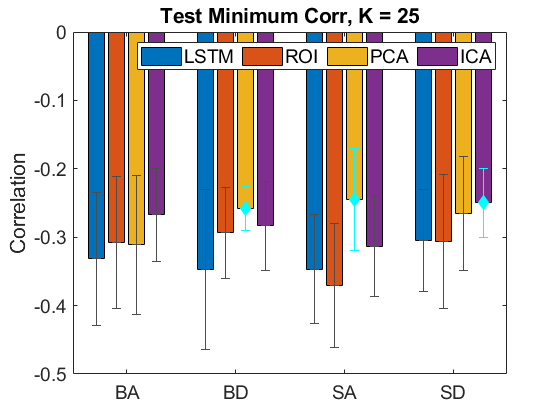}}\negthickspace{}\negthickspace{}\subfloat[]{\includegraphics[width=0.36\columnwidth]{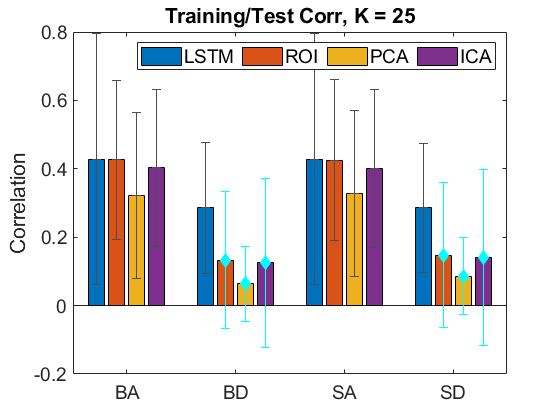}}
\par\end{centering}

\noindent \begin{centering}
\subfloat[]{\includegraphics[width=0.36\columnwidth]{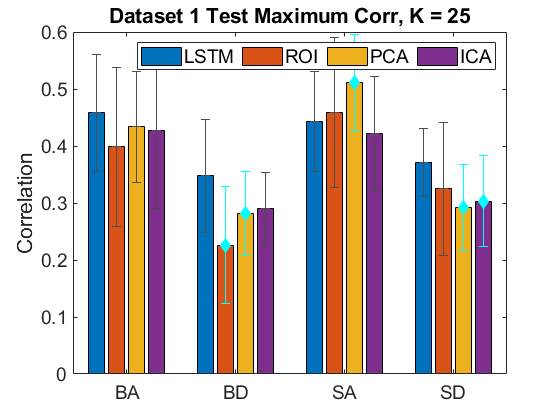}}\subfloat[]{\includegraphics[width=0.36\columnwidth]{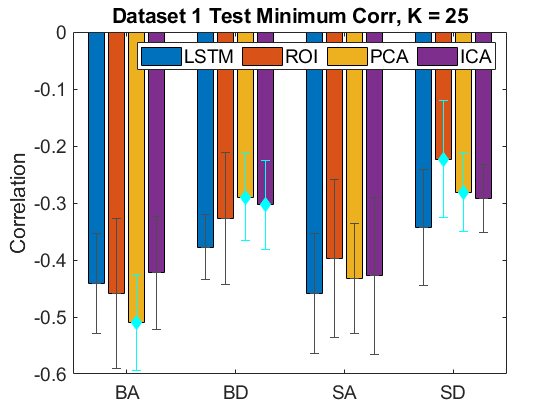}}
\par\end{centering}

\caption{\label{fig:dataset2_25}Results for Dataset 2, with $K=25$ networks.
See Fig.~1 caption for legend.}
\end{figure}
\begin{figure}
\noindent \begin{centering}
\negthinspace{}\negthinspace{}\subfloat[]{\includegraphics[width=0.36\columnwidth]{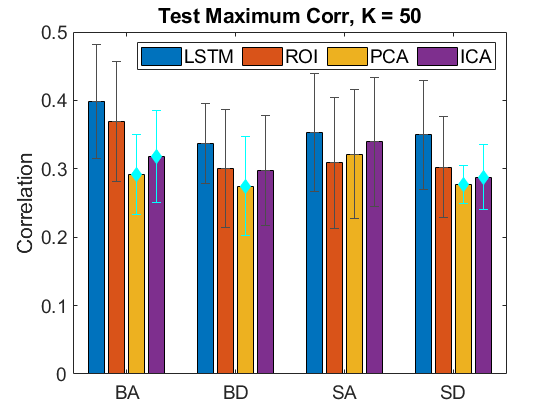}}\negthickspace{}\negthickspace{}\subfloat[]{\includegraphics[width=0.36\columnwidth]{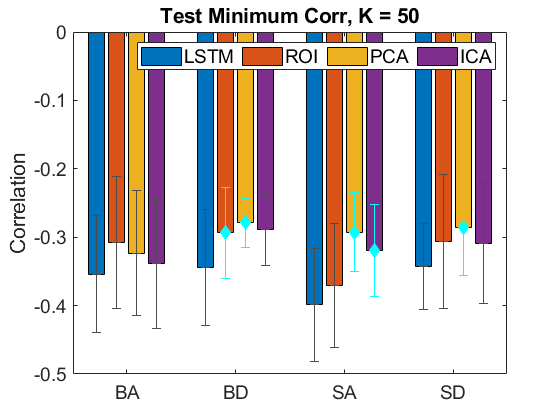}}\negthickspace{}\negthickspace{}\subfloat[]{\includegraphics[width=0.36\columnwidth]{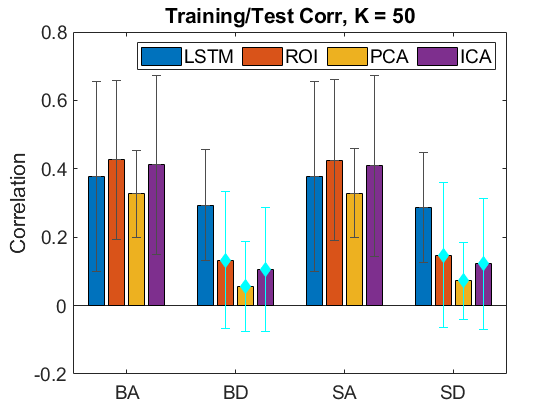}}
\par\end{centering}

\noindent \begin{centering}
\subfloat[]{\includegraphics[width=0.36\columnwidth]{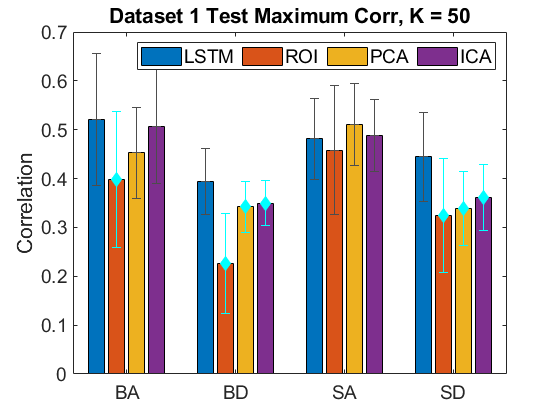}}\subfloat[]{\includegraphics[width=0.36\columnwidth]{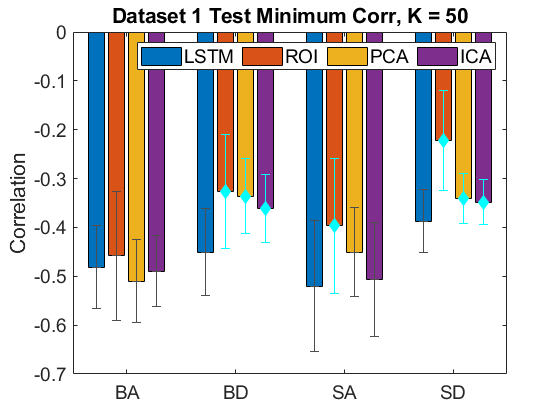}}
\par\end{centering}

\caption{\label{fig:dataset2_50}Results for Dataset 2, with $K=50$ networks.
See Fig.~1 caption for legend.}
\end{figure}

Results for Dataset 1 are shown in Figs.~\ref{fig:dataset1_25} and
\ref{fig:dataset1_50}, and results for Dataset 2 are shown in Figs.~\ref{fig:dataset2_25}
and \ref{fig:dataset2_50}. Cyan markers denote statistically significantly
different results compared to our LSTM method.

Similar trends can be observed across datasets and different numbers
of functional networks to be estimated $K$. Correlation of functional
network activity for subjects in the test set with the design signals
are generally stronger or similar using our LSTM method compared to
other approaches (subfigures (a) and (b)). The correlation between
the values computed from the training set and the test set are clearly
stronger using our LSTM method, particularly for design signals for
dynamic changes in biological and scrambled motion tasks (subfigures
(c)). Together this suggests that our approach estimates functional
networks whose activity patterns (defined by the correlation vectors
$c_{BA}$, $c_{BD}$, $c_{SA}$, $c_{SD}$) are more closely replicated
across subjects within the same dataset compared to the other methods.
Finally, when using functional networks defined by one dataset to
estimate functional network activity in the other dataset, we again
saw that our approach overall resulted in the larger correlations
with the task design signals (subfigures (d) and (e)). This demonstrates
that our LSTM approach is able to find more reliable networks that
are highly associated with task dynamics across different datasets.

\section{Conclusions}

We have presented a method for determining more reproducible functional
networks whose activity is more strongly associated with a given task
paradigm. Our approach uses unsupervised training of LSTMs to learn
to generate the fMRI ROI time-series. We demonstrated stronger correlations
between the activity of the LSTM-derived functional networks with
the design signals for a biological motion perception task. These
results translated better across subjects within the same dataset
and across datasets. This suggests the networks found are more reproducible
and more reliably characterize the network activity in the brain,
which is essential for better characterizing the neural correlates
of a target task. 

{\small{}\bibliographystyle{../ISBI2018/IEEEbib}
\bibliography{isbi2019_ref}
}{\small \par}
\end{document}